\documentclass[letter]{aa} 

\usepackage{graphicx}
\usepackage{comment}
\usepackage{txfonts}
\usepackage{booktabs}
\usepackage[allcolors=blue]{hyperref}
\newcommand{\orcidlink}[1]{\protect\href{https://orcid.org/#1}{\protect\includegraphics[width=8pt]{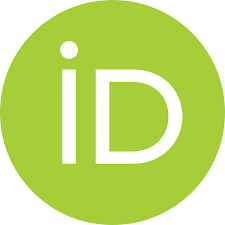}}}

\begin{document}

   \title{Eating planets makes you younger: The magnetic dynamo rejuvenation of GJ 504 by planetary engulfment}

   \subtitle{}
   \titlerunning{The rejuvenated dynamo of GJ~504}

   \author{S. Bellotti \inst{1,2}\orcidlink{0000-0002-2558-6920}
   \and
           C. Pezzotti \inst{3,4}
    \and
           G. Buldgen \inst{3}
    \and
           A. A. Vidotto \inst{1}\orcidlink{0000-0001-5371-2675}
    \and
           D. Evensberget \inst{1,5}\orcidlink{0000-0001-7810-8028}
    \and   
           E. Magaudda\inst{6}\orcidlink{0000-0002-9107-1124}
          }
   \authorrunning{Bellotti et al.}
    
   \institute{
            Leiden Observatory, Leiden University,
            PO Box 9513, 2300 RA Leiden, The Netherlands\\
            \email{bellotti@strw.leidenuniv.nl}
        \and
            Institut de Recherche en Astrophysique et Plan\'etologie,
            Universit\'e de Toulouse, CNRS, IRAP/UMR 5277,
            14 avenue Edouard Belin, F-31400, Toulouse, France
        \and 
        STAR Institute, Université de Liège, Liège, Belgium
        \and
        Istituto Nazionale di Astrofisica – Osservatorio Astronomico di Roma, Via Frascati 33, I-00040, Monteporzio Catone, Italy
        \and Centre for Planetary Habitability (PHAB), Department for Geosciences, University of Oslo, Oslo, Norway
        \and Institut f\"ur Astronomie und Astrophysik, Eberhard-Karls Universit\"at T\"ubingen, Sand 1, 72076 T\"ubingen, Germany
             }

   \date{Received ; accepted }

 
  \abstract
   {With the discovery of a few thousand exoplanets, questions have been raised regarding star-planet interactions and whether the presence of a companion may affect stellar properties. GJ~504 is an excellent target in this context. This evolved ($\rm\sim2\,Gyr$) Sun-like star has a short rotation period (3.4~d) and an intense magnetic activity level, as seen by the X-ray luminosity and the chromospheric diagnostics, which is in stark contrast with what would be expected at such an evolutionary stage.}
   {One possible explanation is that a close-in, Jupiter-mass planet was pushed starwards by the action of stellar tides, inducing a stellar spin-up and ultimately a rejuvenation of the stellar magnetic dynamo. By characterising the large-scale magnetic field and magnetised wind of GJ~504, we aim to provide additional observational constraints to test such scenario.}
   {We analysed spectropolarimetric observations of GJ~504 collected with ESPaDOnS. Using Zeeman-Doppler imaging, we found a large-scale, dipolar, non-axisymmetric magnetic field with an average strength of 5.3\,G, similar to that of evolved early-G type stars. We fed the magnetic field information into our 3D magnetohydrodynamical simulation of the stellar wind and space environment of GJ~504, from which we constrained the wind-driven angular momentum loss ($\rm \dot{J}$). We then compared $\rm \dot{J}$ to rotational evolutionary tracks of GJ~504 for two scenarios: evolution with and without the engulfment of a close-in, Jupiter-mass companion.}
   {Between the two scenarios, only the planetary engulfment can explain the observational constraints obtained previously in the literature, such as the stellar rotation and X-ray luminosity, and the $\rm \dot{J}$ we derived and rescaled to account for underestimated magnetic field strength. Although there are many other stars with similar masses and rotation periods whose rotation evolution does not require planet engulfment, we also identified HD~75332 as a second candidate for planet engulfment, suggesting that GJ~504 may not be an isolated case.}
   {}

   \keywords{Stars: magnetic field --
                Stars: activity --
                Stars: evolution --
                Stars: winds --
                Techniques: polarimetric
               }

   \maketitle
%

\section{Introduction}

Currently, 6042 exoplanets have been confirmed with various techniques \citep[Nasa Exoplanet Archive\footnote{As of November 2025, https://exoplanetarchive.ipac.caltech.edu/}][]{Christiansen2025}. The precise characterisation of the exoplanet host stars is essential to fully understand the origin and history of these systems \citep[e.g.][]{Danielski2022}. On the one hand, stars govern and shape the evolution of planetary systems, by determining the conditions for planetary formation, orbital migration, atmospheric erosion, and, notably, habitability. On the other hand, the fundamental properties of the host stars can be significantly influenced by the presence of close-by planets, leading to mutual star-planet interactions (SPIs) of various types \citep[tidal, radiative, magnetic, and wind; see][for a recent review]{Vidotto2025}. 

An interesting target for studying SPIs is GJ~504, an isolated solar-like star (spectral type G0) with a mass of $1.22~M_{\odot}$, hosting a directly imaged substellar companion at a projected distance of $\rm \sim 43~AU$ \citep{Kuzuhara2013}. The evolutionary state of GJ~504 is highly uncertain, with age estimations varying from hundreds of millions of years to several gigayears. Previous studies on GJ~504 showed that the ages derived from spectroscopic analyses \citep[$\rm 4.5^{+2}_{-1}~Gyr$][]{Fuhrmann2015} and isochrone fitting \citep[$\rm 2.5^{+1.0}_{-0.7}~Gyr$][]{dOrazi2017} are incompatible with the one based on activity indicators and gyrochronology \citep[$\rm \sim 160-300~Myr$][e.g.]{Kuzuhara2013}, despite the large uncertainty. Moreover, the study of \citet{Bonnefoy2018} based on direct imaging and interferometry found plausible isochronal ages to be $\rm 21\pm2~Myr$ and $\rm 4.0\pm1.8~Gyr$.

As studied in \citet{Pezzotti2025}, two possible scenarios can be invoked to reconcile the controversial findings of GJ~504's age. First, the rotational evolution of GJ~504 may have featured weakened magnetic braking \citep[e.g.][]{VanSaders2016}, during which the stellar large-scale magnetic field would not efficiently brake the stellar rotation. This is consistent with the rotation period value of 3.4\,d. Second, the star possibly engulfed a close-by ($\rm\lesssim1~AU$) sub-stellar ($\rm\lesssim 3~M_{Jup}$) companion, enhancing its activity \citep{Fuhrmann2015,dOrazi2017}. This tidally driven SPI event would increase the stellar rotation, leading to an enhanced X-ray luminosity and magnetic activity, overall rejuvenating the dynamo process sustaining the stellar magnetic field. The evolutionary models for stellar rotation, Rossby number (that is, the ratio between rotation period and convective turnover time), and X-ray luminosity of \citet{Pezzotti2025} corroborated that a planetary engulfment is a viable scenario and indicated an age of around 2~Gyr. The recent work of \citet{Lazovik2026} shows that the dissipation of internal gravity waves within stars is affected by tidal SPIs. Specifically for GJ~504, a planetary engulfment can explain the short rotation period of the star.

In this letter, we follow up on the work of \citet{Pezzotti2025} on GJ~504 and test whether the stellar wind-driven angular momentum loss rate could be used as an additional constraint to discriminate between one of these two scenarios. We analysed spectropolarimetric observations of GJ~504 in order to reconstruct the stellar large-scale magnetic field and perform magnetohydrodynamical simulations of the stellar wind. We contextualised these new observational constraints with the theoretical evolution tracks of various quantities such as stellar rotation, X-ray luminosity, angular momentum loss, and total angular momentum. Such a comparison allowed us to further discriminate whether GJ~504 evolved without a planet or with a close-in, Jupiter mass planet that was engulfed.

This letter is structured as follows. We outline the large-scale magnetic field reconstruction in Sect.~\ref{sec:zdi} and then describe the 3D magnetohydrodynamical simulations of the GJ~504's stellar wind in Sect.~\ref{sec:wind}. The comparison between observational constraints and stellar evolutionary models is described in Sect.~\ref{sec:rot_models}. We finally draw our conclusions in Sect.\ref{sec:conclusions}.

\section{Magnetic imaging}\label{sec:zdi}

We observed GJ~504 with ESPaDOnS in April 2025. We provide the fundamental properties of GJ~504 in Table~\ref{tab:star_properties} and a description of the observations in Appendix~\ref{app:obs}. We characterised the large-scale component of GJ~504's magnetic field from circularly polarised spectra by means of Zeeman-Doppler imaging \citep[ZDI; for more information see][]{Semel1989,Donati1997}. We employed the \texttt{zdipy} code described in \citet{Folsom2018} and adopted the weak-field approximation, for which Stokes~$V$ is proportional to the derivative of Stokes~$I$ with respect to wavelength \citep[e.g.][]{Landi1992}. As outlined in \citet{Folsom2018}, the local unpolarised line profiles are modelled with a Voigt kernel. We performed a $\chi^2_r$ minimisation between the median of the observed Stokes~$I$ LSD profiles and its model to find the best-fitting parameters of the kernel. We found the optimal values for depth, Gaussian width, and Lorentzian width to be 0.8, 1.63~km~s$^{-1}$, and 1.7~km~s$^{-1}$.

\begin{table}[!t]
\caption{Properties of GJ~504.} 
\label{tab:star_properties}     
\centering                       
\begin{tabular}{l r}    
\toprule
Property & Value\\
\midrule
Spectral Type & G0V$^a$ \\
$V$  [mag] & 5.22$^b$ \\
Distance [pc] & 17.6$^c$ \\ 
T$_\mathrm{eff}$ [K] & $6205\pm20^b$ \\
$\log g$ [cgs] & $4.29\pm0.07^b$ \\
M/H [dex] & $0.22\pm0.04^b$ \\
Mass [M$_\odot$] & $1.29\pm0.05^d$ \\
Radius [R$_\odot$] & $1.35\pm0.04^e$\\
Age [Gyr] & $2.11 \pm 0.46^d$ \\
Ro & $0.62^d$ \\
P$_\mathrm{rot}$ [d] & 3.4$^b$ \\
$v_\mathrm{eq}\sin i$ [km s$^{-1}$] & $6.3\pm1.0^b$ \\
$i$ [$^\circ$] & 18$^b$\\
\bottomrule 
\end{tabular}
\tablefoot{The listed properties are: identifier, spectral type, $V$ band magnitude, distance, effective temperature, surface gravity, metallicity, stellar mass, stellar radius, stellar age, Rossby number, rotation period, equatorial projected velocity, and inclination. The references are: $a)$ \citet{Gray2001}, $b)$ \citet{dOrazi2017}, $c)$ \citet{Gaia2020}, $d)$ \citet{Pezzotti2025}, and $e)$ \citet{diMauro2022}.}
\end{table}

For the ZDI input parameters, we used the values of the stellar rotation period (P$_\mathrm{rot}=3.4$~d), projected equatorial velocity ($v_\mathrm{eq}\sin(i)=6.3$~km~s$^{-1}$), and inclination ($i=18^{\circ}$) as listed in Table~\ref{tab:star_properties}. We set the maximum degree of spherical harmonic coefficients, $\ell_\mathrm{max}=10$, but a lower value could have been used without changing the results as most of the magnetic energy is stored in the low-$\ell$ degrees. We set the limb darkening coefficient to 0.6 \citep{Claret2011}. Finally, the differential rotation search was inconclusive; hence, we assumed solid body rotation.

The Stokes~$V$ LSD profiles and their ZDI fits are shown in Fig.~\ref{fig:stokesV}. The reconstructed ZDI map is shown in Fig.~\ref{fig:zdi_map}. The magnetic field has a mean strength of $\rm \langle |B_V| \rangle=5.3\,G$ and a maximum of $\rm |B_{max}|=12.9\,G$. The magnetic energy is $\rm \langle B^2\rangle=0.35\times10^{2}G^2$ and 88\% of it is stored in the poloidal component. Of this component, 61\% is stored in the dipolar component, 28\% in the quadrupolar component, and 9\% in the octupolar component. The field is mostly non-axisymmetric, since only 25\% of the magnetic energy is accounted in the $\ell>0$ and $m=0$ modes. Finally, the axisymmetric-poloidal component accounts for 21\% and the axisymmetric-toroidal component for 54\%.

\section{Stellar wind modelling}\label{sec:wind}

\begin{figure}[t]
    \centering
    \includegraphics[width=0.6\columnwidth]{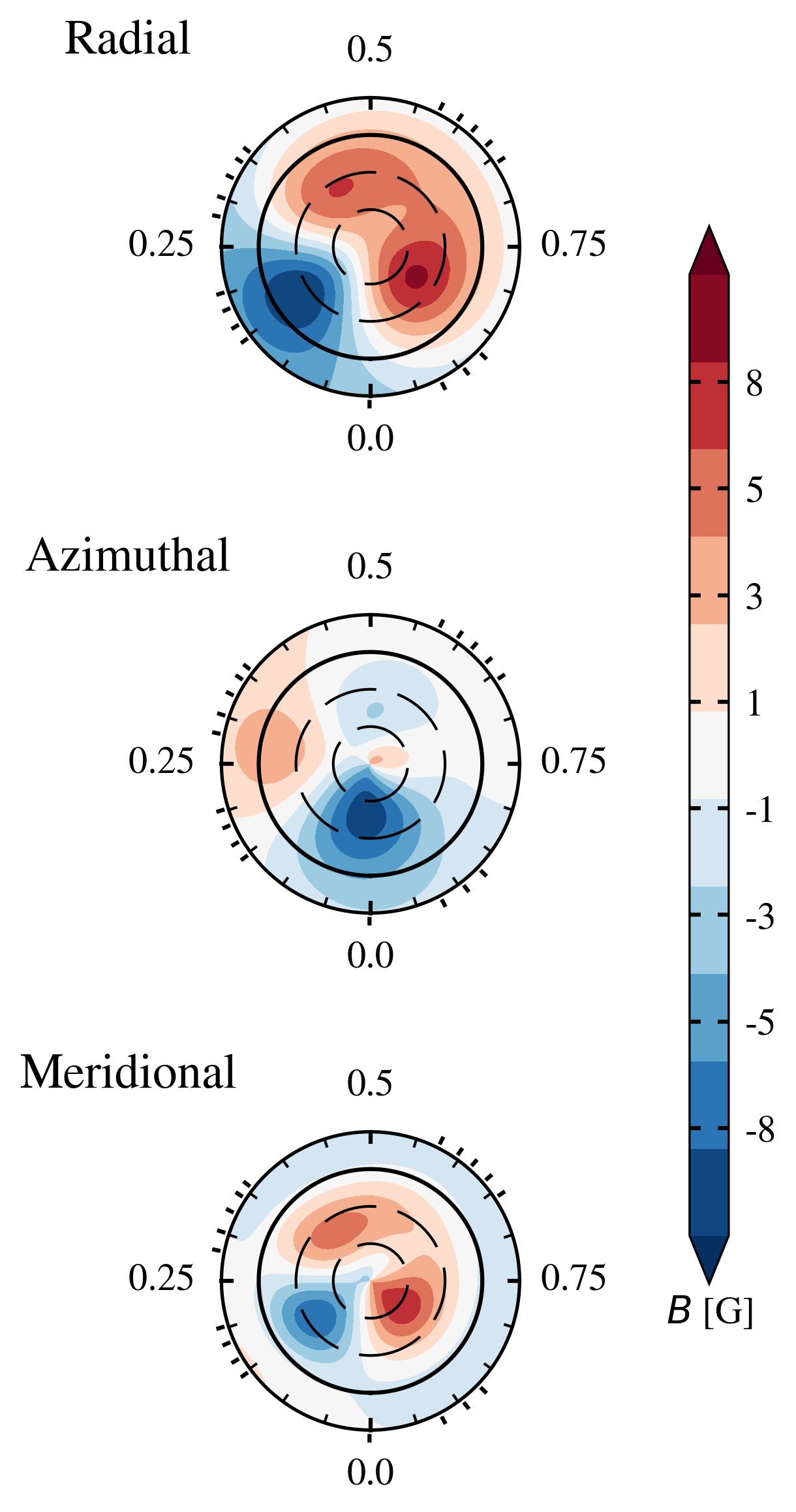}
    \caption{Reconstructed large-scale magnetic field map in flattened polar view. From the left, the radial, azimuthal, and meridional components of the magnetic field vector are illustrated. Concentric circles represent different stellar latitudes: -30\,$^{\circ}$, +30\,$^{\circ}$, and +60\,$^{\circ}$ (dashed lines), as well as the equator (solid line). The radial ticks are located at the rotational phases when the observations were collected (see Table~\ref{tab:log}). The colour indicates the polarity and strength of the magnetic field.}
    \label{fig:zdi_map}
\end{figure}

We simulated the stellar wind of GJ~504 using the Space Weather Modelling Framework \citep[\texttt{SWMF},\ ][]{Toth2012} and specifically the Alfvén wave solar model \citep[\texttt{AWSoM},\ ][]{vanderHolst2014} using the ZDI reconstructed map described in Sect~\ref{sec:zdi} at the inner boundary of the model. A detailed description of the methodology behind the wind models can be found in the recent works of, for instance, \citet{oFionnagain2019} and \citet{Alvarado-Gomez2022b}. The models fix the radial magnetic field component at the inner boundary to the ZDI-derived values (see Fig.~\ref{fig:zdi_map}), while the transverse components are left to evolve as the numerical solution relaxes towards steady state. Except for the stellar mass, radius, and rotation period, the other parameters used in the model are the same as those used for the solar wind and in our previous models \citep[e.g.][]{Evensberget2023}.

\begin{figure}[t]
    \centering
    \includegraphics[width=0.9\columnwidth]{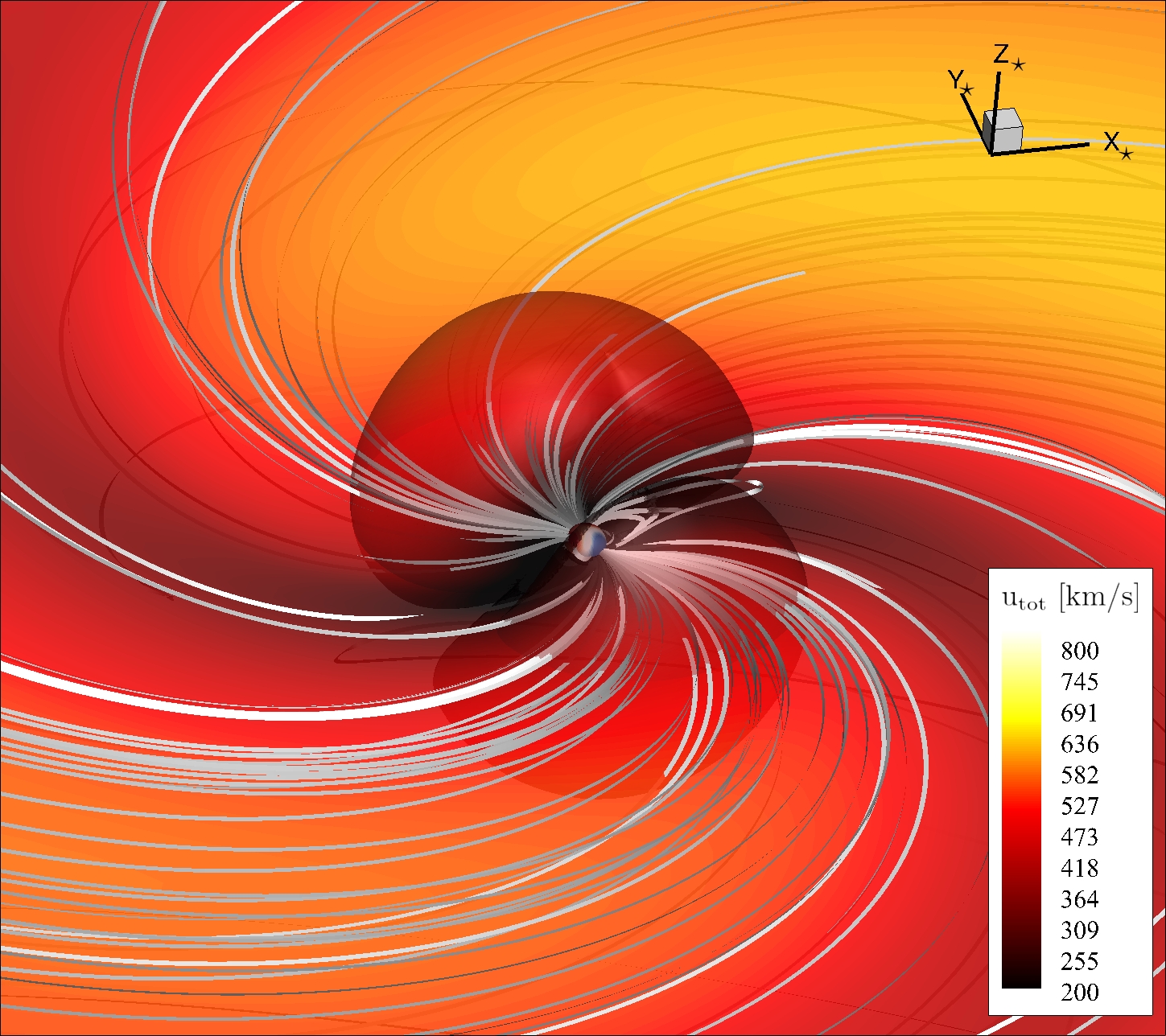}
    \caption{Simulated stellar wind of GJ~504. The star is at the centre and its rotation axis lies along the positive $\rm z_\star$. The $\rm x_\star-y_\star$ plane is coloured by the total wind velocity. The Alfv\'en surface is the region of space where the local wind speed matches the Alfv\'en wave speed, $\rm v_A B_w / \sqrt{4\pi\rho_w}$ (in cgs units), and is depicted as a translucent surface with two lobes, as expected for stars with dominant dipolar large-scale field configurations. The colour bar indicates the total wind velocity ($\rm u_{tot}$).}
    \label{fig:3d_wind}
\end{figure}

In Fig.~\ref{fig:3d_wind}, we present the steady-state output of our simulations centred on the star. The wind speed (${\bf u}$) increases with the distance from the star, while the local wind density ($\rho_w$) and magnetic field ($\rm B_w$) decrease. Furthermore, the wind exhibits a spiral shape owing to stellar rotation. The white streamlines indicate the magnetic field embedded in the stellar wind: open field lines are located at the stellar magnetic poles and closed field lines are visible in the regions where the magnetic polarity switches over. 

We computed the mass loss rate of the wind ($\rm \dot{M}$) by integrating the mass flux over a closed spherical surface ($\Sigma$) centred on the star following Eq.~8 in \citet{Vidotto2014}. We estimated $\rm \dot{M} = 1.07\times10^{-13}\,M_\odot/yr$, which is about 5.4 times larger than the solar wind-mass loss rate. We note that $\rm \dot{M}$ should not vary with the choice of surface, $\Sigma$, as long as such surface encloses the star. We found only a variation of 0.6\% by varying the spherical surface radius across the simulation domain. We then computed the angular momentum loss rate ($\rm \dot{J}$), which regulates the spin-down of the star with age, using Eq.~9 in \citet{Vidotto2014}. We estimated $\rm \dot{J} = 7.53\times10^{31}\,erg$ , which is conserved at most within 5\%. This value is a factor of 23 larger than the average mass loss rate of the Sun computed for cycle 23-24 \citep{Finley2019}.  

\section{Rotational evolution models}\label{sec:rot_models}

We followed \citet{Pezzotti2025} and studied two evolutionary scenarios for GJ~504 (see Appendix~\ref{app:rot_model} for the model setup): one without tidally interacting Jupiter-mass companions and one in which the star was orbited by a Jupiter-mass, close-in planet with an initial orbital distance of $\rm a_{in} = 0.025~AU$ at the moment of dispersal of the protoplanetary disc. The planet subsequently migrated starward due to tidal interactions. \citet{Pezzotti2025} found that the engulfment scenario agrees better with the short rotational period and bright X-ray luminosity of the star. 

\begin{figure*}
    \centering
\includegraphics[width=\textwidth]{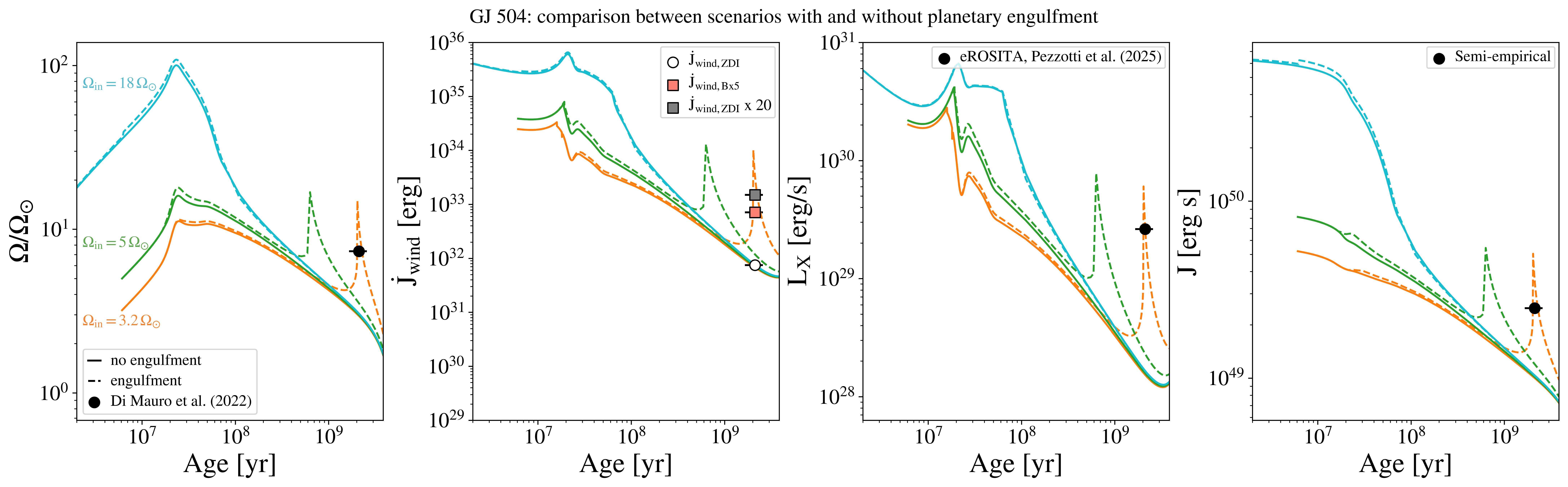} 
    \caption{Evolutionary tracks of GJ~504. The scenarios without (solid lines) and with engulfment (dashed lines) are shown. The spikes in the tracks correspond to the maximum transfer of angular momentum from the planetary orbit to the star, that occurs when the planet reaches the Roche limit. From the left: the stellar surface rotation rate, with the observational constraint taken from \citet{diMauro2022}; stellar wind-driven angular momentum loss tracks with the unscaled $\rm \dot{J}_{wind}$ derived in Sect.~\ref{sec:wind} (white marker), the $\rm \dot{J}_{wind, B\times5}$ obtained by scaling the magnetic field strength (orange marker), and $\rm \dot{J}_{wind,ZDI}$ rescaled by a factor of 20 (gray marker); X-ray luminosity, with the black dot indicating the observational constraint of \citet{Pezzotti2025}; and global angular momentum, with the black dot showing the semi-empirical angular momentum, which is the model momentum of inertia multiplied by the surface rotation rate from \citet{diMauro2022}.}
    \label{Fig:single_all}
\end{figure*}

Our results are shown in Fig.~\ref{Fig:single_all}, where we compare the evolutionary tracks of the normalised stellar rotation rate, $\rm \Omega/\Omega_{\odot}$ (with $\rm \Omega_{\odot} = 2.9 \times 10^{-6}~Hz$), the angular momentum loss, $\rm  \dot{J}_{wind}$, the X-ray luminosity, $\rm L_X$, and the angular momentum, $\rm J$. The tracks were computed with initial values of $\rm \Omega_{in}(\Omega_{\odot}) = 3.2, 5, 18$, which were chosen to reproduce the 25th, 50th, and 90th rotational percentiles in open clusters and stellar associations \citep[e.g.][]{Eggenberger2019a}. 

In the $\rm \dot{J}_{wind}$ panel, we include the angular momentum loss rate derived in Sect.~\ref{sec:wind} ($\rm \dot{J}_{wind,ZDI}$). This value alone is in agreement with the tracks of isolated evolution (solid lines), but these same tracks fail to reproduce the observed rotation period, X-ray luminosity, and semi-empirical global angular momentum of the star. When the value $\rm \dot{J}_{wind,ZDI}$ is rescaled by a constant factor, motivated by the fact that ZDI is known to underestimate the magnetic field strength \citep{Yadav2015,Lehmann2019}, it becomes compatible with the planetary engulfment tracks. Specifically, the evolutionary tracks including engulfment (for $\Omega$ between 3.2 and $\rm 4.0~\Omega_\odot$) match all the four independent observational constraints. A more detailed explanation on the scaling factors can be found in Appendix~\ref{app:torque_scaling}.

\section{Conclusions}\label{sec:conclusions}

We followed up on the recent work of \citet{Pezzotti2025} on GJ~504, a Sun-like star that has a controversial evolutionary state. \citet{Pezzotti2025} compared the rotation and X-ray luminosity of GJ~504 derived from stellar evolution models to observations. To reconcile the observations with the theoretical tracks at the age of the star, the authors proposed a scenario in which GJ~504 engulfed a close-in, Jupiter-mass planet at an age of $\rm \sim 1-2$ Gyr. This tidally induced SPI may have spun up the star and in turn rejuvenated its dynamo, leading to enhanced magnetic activity. With this work, we analysed the large-scale magnetic field and magnetised wind of GJ~504 and found additional constraints that support the planetary engulfment scenario.

We compared the observational constraint on the stellar angular momentum loss to the evolutionary tracks of GJ~504 obtained from a scenario in which the star evolved alone and another in which the star engulfed an exoplanet. If the star is assumed to evolve in isolation, the computed $\dot{J}$ is the only quantity that matches the evolution. However, if $\dot{J}$ is rescaled to counteract the fact that Zeeman Doppler imaging may underestimate the magnetic field strength, $\dot{J}$ becomes incompatible with the star evolving in isolation. When considering the engulfment of a planet instead, all observational constraints (stellar rotation, X-ray luminosity, total angular momentum, and rescaled $\dot{J}$) are consistent with the evolutionary tracks. Ultimately, this provides further support of the planetary engulfment scenario proposed by \citet{Pezzotti2025}.

To understand whether GJ~504 is an isolated case, we performed a similar analysis for HD~75332, which is another Sun-like star lying in the vicinity of GJ~504 in terms of mass, rotation period, and magnetic field properties (see Appendix~\ref{app:HD75332}). We found analogous results to GJ~504, although other stars with similar properties do not show this anomaly \citep{Pezzotti2025b}. Additional observations from a dedicated campaign are required to provide more constraints on these phenomena, such as radial velocity follow-ups to study oscillations and constrain the age of GJ~504. In general, from both spectroscopic studies of chemical enrichment \citep[e.g.][]{Soares2025} and tidal SPI studies \citep{Lazovik2026}, an occurrence rate of 20\% for planetary engulfments with detectable signatures for Sun-like stars is expected.

\begin{acknowledgements}
SB and AAV acknowledge funding by the Dutch Research Council (NWO) under the project "Exo-space weather and contemporaneous signatures of star-planet interactions" (with project number OCENW.M.22.215 of the research programme "Open Competition Domain Science- M"). AAV and DE acknowledge funding from the European Research Council (ERC) under the European Union's Horizon 2020 research and innovation programme (grant agreement No 817540, ASTROFLOW). AAV acknowledges funding from the Dutch Research Council (NWO), with project number VI.C.232.041 of the Talent Programme Vici. GB acknowledges funding from the Fonds National de la Recherche Scientifique (FNRS) as a postdoctoral researcher. CP thanks the Belgian Federal Science Policy Office (BELSPO) for the financial support in the framework of the PRODEX Program of the European Space Agency (ESA) under contract number 4000141194. EM is supported by the Deutsche Forschungsgemeinschaft under grant STE 1068/8. This work used the Dutch national e-infrastructure with the support of the SURF Cooperative using grant nos. EINF-2218 and EINF-5173. Based on observations obtained at the CFHT which is operated by the National Research Council of Canada, the Institut National des Sciences de l'Univers of the Centre National de la Recherche Scientique of France, and the University of Hawaii.

\end{acknowledgements}

%
%

\bibliographystyle{aa}
\bibliography{biblio}

\begin{thebibliography}{50}
\expandafter\ifx\csname natexlab\endcsname\relax\def\natexlab#1{#1}\fi

\bibitem[{{Alvarado-G{\'o}mez} {et~al.}(2022){Alvarado-G{\'o}mez}, {Cohen}, {Drake}, {Fraschetti}, {Poppenhaeger}, {Garraffo}, {Chebly}, {Ilin}, {Harbach}, \& {Kochukhov}}]{Alvarado-Gomez2022b}
{Alvarado-G{\'o}mez}, J.~D., {Cohen}, O., {Drake}, J.~J., {et~al.} 2022, \apj, 928, 147

\bibitem[{{Bonnefoy} {et~al.}(2018){Bonnefoy}, {Perraut}, {Lagrange}, {Delorme}, {Vigan}, {Line}, {Rodet}, {Ginski}, {Mourard}, {Marleau}, {Samland}, {Tremblin}, {Ligi}, {Cantalloube}, {Molli{\`e}re}, {Charnay}, {Kuzuhara}, {Janson}, {Morley}, {Homeier}, {D'Orazi}, {Klahr}, {Mordasini}, {Lavie}, {Baudino}, {Beust}, {Peretti}, {Musso Bartucci}, {Mesa}, {B{\'e}zard}, {Boccaletti}, {Galicher}, {Hagelberg}, {Desidera}, {Biller}, {Maire}, {Allard}, {Borgniet}, {Lannier}, {Meunier}, {Desort}, {Alecian}, {Chauvin}, {Langlois}, {Henning}, {Mugnier}, {Mouillet}, {Gratton}, {Brandt}, {Mc Elwain}, {Beuzit}, {Tamura}, {Hori}, {Brandner}, {Buenzli}, {Cheetham}, {Cudel}, {Feldt}, {Kasper}, {Keppler}, {Kopytova}, {Meyer}, {Perrot}, {Rouan}, {Salter}, {Schmidt}, {Sissa}, {Zurlo}, {Wildi}, {Blanchard}, {De Caprio}, {Delboulb{\'e}}, {Maurel}, {Moulin}, {Pavlov}, {Rabou}, {Ramos}, {Roelfsema}, {Rousset}, {Stadler}, {Rigal}, \& {Weber}}]{Bonnefoy2018}
{Bonnefoy}, M., {Perraut}, K., {Lagrange}, A.~M., {et~al.} 2018, \aap, 618, A63

\bibitem[{{Brown} {et~al.}(2021){Brown}, {Marsden}, {Mengel}, {Jeffers}, {Millburn}, {Mittag}, {Petit}, {Vidotto}, {Morin}, {See}, {Jardine}, {Gonz{\'a}lez-P{\'e}rez}, {Gonz{\'a}lez-P{\'e}rez}, \& {BCool Collaboration}}]{Brown2021}
{Brown}, E.~L., {Marsden}, S.~C., {Mengel}, M.~W., {et~al.} 2021, \mnras, 501, 3981

\bibitem[{{Buldgen} {et~al.}(2019){Buldgen}, {Farnir}, {Pezzotti}, {Eggenberger}, {Salmon}, {Montalban}, {Ferguson}, {Khan}, {Bourrier}, {Rendle}, {Meynet}, {Miglio}, \& {Noels}}]{Buldgen2019}
{Buldgen}, G., {Farnir}, M., {Pezzotti}, C., {et~al.} 2019, \aap, 630, A126

\bibitem[{{Butler} {et~al.}(1997){Butler}, {Marcy}, {Williams}, {Hauser}, \& {Shirts}}]{Butler1997}
{Butler}, R.~P., {Marcy}, G.~W., {Williams}, E., {Hauser}, H., \& {Shirts}, P. 1997, \apjl, 474, L115

\bibitem[{{Christiansen} {et~al.}(2025){Christiansen}, {McElroy}, {Harbut}, {Ciardi}, {Crane}, {Good}, {Hardegree-Ullman}, {Kesseli}, {Lund}, {Lynn}, {Muthiar}, {Nilsson}, {Oluyide}, {Papin}, {Rivera}, {Swain}, {Susemiehl}, {Tam}, {van Eyken}, \& {Beichman}}]{Christiansen2025}
{Christiansen}, J.~L., {McElroy}, D.~L., {Harbut}, M., {et~al.} 2025, PSJ, 6, 186

\bibitem[{{Claret} \& {Bloemen}(2011)}]{Claret2011}
{Claret}, A. \& {Bloemen}, S. 2011, A\&A, 529, A75

\bibitem[{{Danielski} {et~al.}(2022){Danielski}, {Brucalassi}, {Benatti}, {Campante}, {Delgado-Mena}, {Rainer}, {Sacco}, {Adibekyan}, {Biazzo}, {Bossini}, {Bruno}, {Casali}, {Kabath}, {Magrini}, {Micela}, {Morello}, {Palladino}, {Sanna}, {Sarkar}, {Sousa}, {Tsantaki}, {Turrini}, \& {Van der Swaelmen}}]{Danielski2022}
{Danielski}, C., {Brucalassi}, A., {Benatti}, S., {et~al.} 2022, Experimental Astronomy, 53, 473

\bibitem[{{Di Mauro} {et~al.}(2022){Di Mauro}, {Reda}, {Mathur}, {Garc{\'\i}a}, {Buzasi}, {Corsaro}, {Benomar}, {Gonz{\'a}lez Cuesta}, {Stassun}, {Benatti}, {D'Orazi}, {Giovannelli}, {Mesa}, \& {Nardetto}}]{diMauro2022}
{Di Mauro}, M.~P., {Reda}, R., {Mathur}, S., {et~al.} 2022, \apj, 940, 93

\bibitem[{{Donahue} {et~al.}(1996){Donahue}, {Saar}, \& {Baliunas}}]{Donahue1996}
{Donahue}, R.~A., {Saar}, S.~H., \& {Baliunas}, S.~L. 1996, \apj, 466, 384

\bibitem[{{Donati} {et~al.}(1997){Donati}, {Semel}, {Carter}, {Rees}, \& {Collier Cameron}}]{Donati1997}
{Donati}, J.~F., {Semel}, M., {Carter}, B.~D., {Rees}, D.~E., \& {Collier Cameron}, A. 1997, MNRAS, 291, 658

\bibitem[{{D'Orazi} {et~al.}(2017){D'Orazi}, {Desidera}, {Gratton}, {Lanza}, {Messina}, {Andrievsky}, {Korotin}, {Benatti}, {Bonnefoy}, {Covino}, \& {Janson}}]{dOrazi2017}
{D'Orazi}, V., {Desidera}, S., {Gratton}, R.~G., {et~al.} 2017, \aap, 598, A19

\bibitem[{{Eggenberger} {et~al.}(2019){Eggenberger}, {Buldgen}, \& {Salmon}}]{Eggenberger2019a}
{Eggenberger}, P., {Buldgen}, G., \& {Salmon}, S.~J.~A.~J. 2019, \aap, 626, L1

\bibitem[{{Evensberget} {et~al.}(2023){Evensberget}, {Marsden}, {Carter}, {Salmeron}, {Vidotto}, {Folsom}, {Kavanagh}, {Pineda}, {Driessen}, \& {Strickert}}]{Evensberget2023}
{Evensberget}, D., {Marsden}, S.~C., {Carter}, B.~D., {et~al.} 2023, \mnras, 524, 2042

\bibitem[{{Evensberget} \& {Vidotto}(2024)}]{Evensberget2024}
{Evensberget}, D. \& {Vidotto}, A.~A. 2024, \mnras, 529, L140

\bibitem[{{Finley} {et~al.}(2019{\natexlab{a}}){Finley}, {Deshmukh}, {Matt}, {Owens}, \& {Wu}}]{Finley2019}
{Finley}, A.~J., {Deshmukh}, S., {Matt}, S.~P., {Owens}, M., \& {Wu}, C.-J. 2019{\natexlab{a}}, \apj, 883, 67

\bibitem[{{Finley} \& {Matt}(2018)}]{Finley2018}
{Finley}, A.~J. \& {Matt}, S.~P. 2018, \apj, 854, 78

\bibitem[{{Finley} {et~al.}(2019{\natexlab{b}}){Finley}, {See}, \& {Matt}}]{Finley2019a}
{Finley}, A.~J., {See}, V., \& {Matt}, S.~P. 2019{\natexlab{b}}, \apj, 876, 44

\bibitem[{{Folsom} {et~al.}(2018){Folsom}, {Bouvier}, {Petit}, {L{\`e}bre}, {Amard}, {Palacios}, {Morin}, {Donati}, \& {Vidotto}}]{Folsom2018}
{Folsom}, C.~P., {Bouvier}, J., {Petit}, P., {et~al.} 2018, MNRAS, 474, 4956

\bibitem[{Folsom {et~al.}(2025)Folsom, Erba, Petit, Seadrow, Stanley, Natan, Zaire, Oksala, Villadiego~Forero, Moore, \& Catalan~Olais}]{Folsom2025}
Folsom, C.~P., Erba, C., Petit, V., {et~al.} 2025, Journal of Open Source Software, 10, 7891

\bibitem[{{Fuhrmann} \& {Chini}(2015)}]{Fuhrmann2015}
{Fuhrmann}, K. \& {Chini}, R. 2015, \apj, 806, 163

\bibitem[{{Gaia Collaboration}(2020)}]{Gaia2020}
{Gaia Collaboration}. 2020, VizieR Online Data Catalog, I/350

\bibitem[{{Gray} {et~al.}(2001){Gray}, {Napier}, \& {Winkler}}]{Gray2001}
{Gray}, R.~O., {Napier}, M.~G., \& {Winkler}, L.~I. 2001, \aj, 121, 2148

\bibitem[{{Johnstone} {et~al.}(2021){Johnstone}, {Bartel}, \& {G{\"u}del}}]{Johnstone2021}
{Johnstone}, C.~P., {Bartel}, M., \& {G{\"u}del}, M. 2021, \aap, 649, A96

\bibitem[{{Kochukhov} {et~al.}(2010){Kochukhov}, {Makaganiuk}, \& {Piskunov}}]{Kochukhov2010a}
{Kochukhov}, O., {Makaganiuk}, V., \& {Piskunov}, N. 2010, A\&A, 524, A5

\bibitem[{{Kuzuhara} {et~al.}(2013){Kuzuhara}, {Tamura}, {Kudo}, {Janson}, {Kandori}, {Brandt}, {Thalmann}, {Spiegel}, {Biller}, {Carson}, {Hori}, {Suzuki}, {Burrows}, {Henning}, {Turner}, {McElwain}, {Moro-Mart{\'\i}n}, {Suenaga}, {Takahashi}, {Kwon}, {Lucas}, {Abe}, {Brandner}, {Egner}, {Feldt}, {Fujiwara}, {Goto}, {Grady}, {Guyon}, {Hashimoto}, {Hayano}, {Hayashi}, {Hayashi}, {Hodapp}, {Ishii}, {Iye}, {Knapp}, {Matsuo}, {Mayama}, {Miyama}, {Morino}, {Nishikawa}, {Nishimura}, {Kotani}, {Kusakabe}, {Pyo}, {Serabyn}, {Suto}, {Takami}, {Takato}, {Terada}, {Tomono}, {Watanabe}, {Wisniewski}, {Yamada}, {Takami}, \& {Usuda}}]{Kuzuhara2013}
{Kuzuhara}, M., {Tamura}, M., {Kudo}, T., {et~al.} 2013, \apj, 774, 11

\bibitem[{{Landi Degl'Innocenti}(1992)}]{Landi1992}
{Landi Degl'Innocenti}, E. 1992, {Magnetic field measurements.}, ed. F.~{Sanchez}, M.~{Collados}, \& M.~{Vazquez}, 71

\bibitem[{{Lazovik} \& {Barker}(2026)}]{Lazovik2026}
{Lazovik}, Y.~A. \& {Barker}, A.~J. 2026, arXiv e-prints, arXiv:2602.03723

\bibitem[{{Lebreton} \& {Reese}(2020)}]{Lebreton2020}
{Lebreton}, Y. \& {Reese}, D.~R. 2020, \aap, 642, A88

\bibitem[{{Lehmann} {et~al.}(2019){Lehmann}, {Hussain}, {Jardine}, {Mackay}, \& {Vidotto}}]{Lehmann2019}
{Lehmann}, L.~T., {Hussain}, G.~A.~J., {Jardine}, M.~M., {Mackay}, D.~H., \& {Vidotto}, A.~A. 2019, \mnras, 483, 5246

\bibitem[{{Matt} {et~al.}(2015){Matt}, {Brun}, {Baraffe}, {Bouvier}, \& {Chabrier}}]{Matt2015}
{Matt}, S.~P., {Brun}, A.~S., {Baraffe}, I., {Bouvier}, J., \& {Chabrier}, G. 2015, \apjl, 799, L23

\bibitem[{{Matt} {et~al.}(2019){Matt}, {Brun}, {Baraffe}, {Bouvier}, \& {Chabrier}}]{Matt2019}
{Matt}, S.~P., {Brun}, A.~S., {Baraffe}, I., {Bouvier}, J., \& {Chabrier}, G. 2019, \apjl, 870, L27

\bibitem[{{{\'O} Fionnag{\'a}in} {et~al.}(2019){{\'O} Fionnag{\'a}in}, {Vidotto}, {Petit}, {Folsom}, {Jeffers}, {Marsden}, {Morin}, {do Nascimento}, \& {BCool Collaboration}}]{oFionnagain2019}
{{\'O} Fionnag{\'a}in}, D., {Vidotto}, A.~A., {Petit}, P., {et~al.} 2019, \mnras, 483, 873

\bibitem[{{Pezzotti} {et~al.}(2026){Pezzotti}, {B{\'e}trisey}, {Buldgen}, {Gilfanov}, {Bikmaev}, {Sunyaev}, {I{\textcommabelow s}{\i}k}, {Gosset}, \& {Wright}}]{Pezzotti2025b}
{Pezzotti}, C., {B{\'e}trisey}, J., {Buldgen}, G., {et~al.} 2026, \aap, 706, A257

\bibitem[{{Pezzotti} {et~al.}(2025){Pezzotti}, {Buldgen}, {Magaudda}, {Farnir}, {Van Grootel}, {Bellotti}, \& {Poppenhaeger}}]{Pezzotti2025}
{Pezzotti}, C., {Buldgen}, G., {Magaudda}, E., {et~al.} 2025, \aap, 694, A179

\bibitem[{{Pezzotti} {et~al.}(2021){Pezzotti}, {Eggenberger}, {Buldgen}, {Meynet}, {Bourrier}, \& {Mordasini}}]{Pezzotti2021}
{Pezzotti}, C., {Eggenberger}, P., {Buldgen}, G., {et~al.} 2021, \aap, 650, A108

\bibitem[{{Privitera} {et~al.}(2016){Privitera}, {Meynet}, {Eggenberger}, {Vidotto}, {Villaver}, \& {Bianda}}]{Privitera2016b}
{Privitera}, G., {Meynet}, G., {Eggenberger}, P., {et~al.} 2016, \aap, 593, A128

\bibitem[{{Rao} {et~al.}(2018){Rao}, {Meynet}, {Eggenberger}, {Haemmerl{\'e}}, {Privitera}, {Georgy}, {Ekstr{\"o}m}, \& {Mordasini}}]{Rao2018}
{Rao}, S., {Meynet}, G., {Eggenberger}, P., {et~al.} 2018, \aap, 618, A18

\bibitem[{{Rao} {et~al.}(2021){Rao}, {Pezzotti}, {Meynet}, {Eggenberger}, {Buldgen}, {Mordasini}, {Bourrier}, {Ekstr{\"o}m}, \& {Georgy}}]{Rao2021}
{Rao}, S., {Pezzotti}, C., {Meynet}, G., {et~al.} 2021, \aap, 651, A50

\bibitem[{{Ryabchikova} {et~al.}(2015){Ryabchikova}, {Piskunov}, {Kurucz}, {Stempels}, {Heiter}, {Pakhomov}, \& {Barklem}}]{Ryabchikova2015}
{Ryabchikova}, T., {Piskunov}, N., {Kurucz}, R.~L., {et~al.} 2015, Phys. Scr., 90, 054005

\bibitem[{{Scuflaire} {et~al.}(2008){Scuflaire}, {Th{\'e}ado}, {Montalb{\'a}n}, {Miglio}, {Bourge}, {Godart}, {Thoul}, \& {Noels}}]{Scuflaire2008}
{Scuflaire}, R., {Th{\'e}ado}, S., {Montalb{\'a}n}, J., {et~al.} 2008, \apss, 316, 83

\bibitem[{{Semel}(1989)}]{Semel1989}
{Semel}, M. 1989, A\&A, 225, 456

\bibitem[{{Soares} {et~al.}(2025){Soares}, {Adibekyan}, {Mordasini}, {Deal}, {Sousa}, {Delgado-Mena}, {Santos}, \& {Dorn}}]{Soares2025}
{Soares}, B.~M.~T.~B., {Adibekyan}, V., {Mordasini}, C., {et~al.} 2025, \aap, 693, A47

\bibitem[{{T{\'o}th} {et~al.}(2012){T{\'o}th}, {van der Holst}, {Sokolov}, {De Zeeuw}, {Gombosi}, {Fang}, {Manchester}, {Meng}, {Najib}, {Powell}, {Stout}, {Glocer}, {Ma}, \& {Opher}}]{Toth2012}
{T{\'o}th}, G., {van der Holst}, B., {Sokolov}, I.~V., {et~al.} 2012, Journal of Computational Physics, 231, 870

\bibitem[{{van der Holst} {et~al.}(2014){van der Holst}, {Sokolov}, {Meng}, {Jin}, {Manchester}, {T{\'o}th}, \& {Gombosi}}]{vanderHolst2014}
{van der Holst}, B., {Sokolov}, I.~V., {Meng}, X., {et~al.} 2014, The Astrophysical Journal, 782, 81

\bibitem[{{van Saders} {et~al.}(2016){van Saders}, {Ceillier}, {Metcalfe}, {Silva Aguirre}, {Pinsonneault}, {Garc{\'\i}a}, {Mathur}, \& {Davies}}]{VanSaders2016}
{van Saders}, J.~L., {Ceillier}, T., {Metcalfe}, T.~S., {et~al.} 2016, \nat, 529, 181

\bibitem[{{Vidotto}(2025)}]{Vidotto2025}
{Vidotto}, A.~A. 2025, \araa, 63, 299

\bibitem[{{Vidotto} {et~al.}(2014{\natexlab{a}}){Vidotto}, {Gregory}, {Jardine}, {Donati}, {Petit}, {Morin}, {Folsom}, {Bouvier}, {Cameron}, {Hussain}, {Marsden}, {Waite}, {Fares}, {Jeffers}, \& {do Nascimento}}]{Vidotto2014a}
{Vidotto}, A.~A., {Gregory}, S.~G., {Jardine}, M., {et~al.} 2014{\natexlab{a}}, \mnras, 441, 2361

\bibitem[{{Vidotto} {et~al.}(2014{\natexlab{b}}){Vidotto}, {Jardine}, {Morin}, {Donati}, {Opher}, \& {Gombosi}}]{Vidotto2014}
{Vidotto}, A.~A., {Jardine}, M., {Morin}, J., {et~al.} 2014{\natexlab{b}}, MNRAS, 438, 1162

\bibitem[{{Yadav} {et~al.}(2015){Yadav}, {Christensen}, {Morin}, {Gastine}, {Reiners}, {Poppenhaeger}, \& {Wolk}}]{Yadav2015}
{Yadav}, R.~K., {Christensen}, U.~R., {Morin}, J., {et~al.} 2015, ApJl, 813, L31

\end{thebibliography}

\begin{appendix} 

\onecolumn

\section{Journal of observations}\label{app:obs}

We collected 20 ESPaDOnS observations between April 5th and 9th, each with an exposure time of 120\,s per polarimetric sequence. ESPaDOns is the spectropolarimeter on the 3.6~m Canada-France-Hawaii-Telescope (CFHT) located atop Mauna Kea in Hawaii. The instrument has a spectral coverage from 380~nm to 1050~nm, and a median spectral resolving power of 65~000 after data reduction. The observations were carried out in circular polarisation mode, providing both unpolarised (Stokes~$I$) and circularly polarised (Stokes~$V$) high-resolution spectra. The data were reduced with the \texttt{LIBRE-ESPRIT} pipeline \citep{Donati1997}. The first observation on April 5th recorded a significantly lower signal-to-noise ratio (S/N) compared to the other observations, due to the presence of clouds that attenuated the signal. For this reason the observation was discarded from the analyses described in the following sections. The signal-to-noise ratio at 650~nm per resolution element ($2.6~\rm km\,s^{-1}$ velocity bin) per polarimetric sequence of the other observations ranges between 146 and 390, with an average of 329. In the next sections, the observations will be phased with the following ephemeris
\begin{align}
    \mathrm{HJD} = \mathrm{HJD}_0 + \mathrm{P}_\mathrm{rot}\cdot n_\mathrm{cyc},
    \label{eq:ephemeris}
\end{align}
where HJD$_\mathrm{0}$ is the heliocentric Julian Date reference (the first one of the time series for each star, see Table~\ref{tab:log}), P$_\mathrm{rot}$ is the rotation period of the star (see Table~\ref{tab:star_properties}), and $n_\mathrm{cyc}$ represents the rotation cycle.

We applied least-squares deconvolution \citep[LSD][]{Donati1997,Kochukhov2010a} to the Stokes~$I$ and $V$ spectra using the \textsc{lsdpy} code which is part of the Specpolflow software \citep{Folsom2025}\footnote{Available at \href{https://github.com/folsomcp/LSDpy}{https://github.com/folsomcp/LSDpy}}. This is a cross-correlation technique that deconvolves an observed spectrum with a line list to obtain an average, high-S/N line profile. We used the Vienna Atomic Line Database\footnote{\url{http://vald.astro.uu.se/}} \citep[VALD,][]{Ryabchikova2015} to generate a synthetic line list corresponding to a stellar temperature of 6250~K and surface gravity of $\log g=4.5$ (cm~s$^{-2}$). When performing LSD, we excluded spectral regions affected by telluric bands and the H$\alpha$ line: [627,632], [655.5,657], [686,697], [716,734], [759,770], [813,835], and [895,986]~nm. In total, we used 7100 atomic spectral lines and we adopted a normalisation wavelength and Land\'e factor (that is, the magnetic sensitivity of the line indicated as $\rm g_{eff}$) of 700~nm and 1.2 \citep[see][for more details]{Kochukhov2010a}. The output is Stokes~$I$ and $V$ LSD profiles with augmented S/N. We re-normalised the Stokes~$I$ LSD profiles continuum to unity by fitting a linear model to the region outside the line, to include residuals of continuum normalisation at the level of the spectra. The Stokes~$V$ profiles were correspondingly rescaled with the same fit.\\

\begin{sidewaystable*}
\caption{Journal of ESPaDOnS observations for GJ~504.}
\label{tab:log}     
\centering                       
\begin{tabular}{l c c c c c r c c c c c c c c }      
\toprule
Date & UT & HJD & $n_\mathrm{cyc}$ & S/N & $\sigma_\mathrm{LSD}$ & B$_\ell$ & $S$ index & $\log R'_\mathrm{HK}$ & H$\alpha$ index & \ion{Ca}{II} IRT\\
 & [hh:mm:ss] &  & & & [$10^{-5}I_c$] & [G] & & & & \\
\midrule
*Apr 05 & 09:00:26 & 2460770.8805 & 0.00 & 88 & 24.4 & \ldots & \ldots & \ldots & \ldots & \ldots \\
Apr 05 & 09:05:18 & 2460770.8839 & 0.00 & 233 & 8.1 & $12.01\pm4.25$ & $0.361\pm0.132$ & $-4.362\pm0.213$ & $0.320\pm0.003$ & $0.861\pm0.004$ \\
Apr 05 & 10:43:38 & 2460770.9522 & 0.02 & 146 & 14.6 & $10.01\pm7.02$ & $0.414\pm0.266$ & $-4.284\pm0.360$ & $0.320\pm0.004$ & $0.860\pm0.007$ \\
Apr 06 & 07:57:57 & 2460771.8372 & 0.28 & 390 & 5.0 & $2.53\pm2.53$ & $0.344\pm0.066$ & $-4.391\pm0.115$ & $0.319\pm0.002$ & $0.853\pm0.003$ \\
Apr 06 & 09:33:16 & 2460771.9034 & 0.30 & 369 & 5.1 & $1.67\pm2.58$ & $0.353\pm0.067$ & $-4.374\pm0.112$ & $0.319\pm0.002$ & $0.855\pm0.003$ \\
Apr 06 & 11:24:55 & 2460771.9809 & 0.32 & 382 & 5.5 & $4.88\pm2.50$ & $0.354\pm0.066$ & $-4.374\pm0.110$ & $0.319\pm0.002$ & $0.852\pm0.003$ \\
Apr 06 & 12:57:43 & 2460772.0453 & 0.34 & 373 & 5.0 & $2.96\pm2.59$ & $0.355\pm0.069$ & $-4.372\pm0.115$ & $0.320\pm0.002$ & $0.850\pm0.003$ \\
Apr 06 & 14:02:05 & 2460772.0900 & 0.36 & 372 & 4.8 & $5.02\pm2.61$ & $0.333\pm0.072$ & $-4.409\pm0.131$ & $0.319\pm0.002$ & $0.851\pm0.003$ \\
Apr 07 & 07:53:26 & 2460772.8340 & 0.57 & 358 & 5.9 & $3.26\pm2.82$ & $0.365\pm0.077$ & $-4.356\pm0.124$ & $0.320\pm0.002$ & $0.855\pm0.003$ \\
Apr 07 & 09:40:27 & 2460772.9083 & 0.60 & 318 & 6.0 & $5.96\pm3.11$ & $0.345\pm0.088$ & $-4.388\pm0.151$ & $0.319\pm0.002$ & $0.858\pm0.003$ \\
Apr 07 & 11:10:03 & 2460772.9706 & 0.61 & 300 & 6.8 & $6.30\pm3.29$ & $0.352\pm0.094$ & $-4.376\pm0.158$ & $0.319\pm0.002$ & $0.856\pm0.003$ \\
Apr 07 & 12:52:14 & 2460773.0415 & 0.64 & 351 & 6.1 & $5.36\pm2.84$ & $0.345\pm0.080$ & $-4.389\pm0.139$ & $0.320\pm0.002$ & $0.859\pm0.003$ \\
Apr 07 & 14:34:45 & 2460773.1127 & 0.66 & 254 & 8.2 & $15.60\pm4.12$ & $0.346\pm0.149$ & $-4.386\pm0.254$ & $0.319\pm0.003$ & $0.860\pm0.004$ \\
Apr 08 & 08:11:11 & 2460773.8464 & 0.87 & 334 & 5.7 & $3.87\pm2.94$ & $0.374\pm0.083$ & $-4.341\pm0.129$ & $0.322\pm0.002$ & $0.863\pm0.003$ \\
Apr 08 & 10:05:34 & 2460773.9258 & 0.90 & 305 & 6.8 & $-1.14\pm3.30$ & $0.357\pm0.095$ & $-4.368\pm0.156$ & $0.320\pm0.002$ & $0.862\pm0.003$ \\
Apr 08 & 12:19:40 & 2460774.0189 & 0.92 & 369 & 4.6 & $-3.72\pm2.67$ & $0.362\pm0.072$ & $-4.360\pm0.116$ & $0.320\pm0.002$ & $0.858\pm0.003$ \\
Apr 09 & 06:41:40 & 2460774.7842 & 1.15 & 331 & 6.2 & $0.04\pm3.03$ & $0.333\pm0.090$ & $-4.410\pm0.162$ & $0.319\pm0.002$ & $0.856\pm0.003$ \\
Apr 09 & 08:04:19 & 2460774.8416 & 1.17 & 361 & 5.4 & $0.94\pm2.70$ & $0.358\pm0.072$ & $-4.367\pm0.119$ & $0.319\pm0.002$ & $0.854\pm0.003$ \\
Apr 09 & 09:39:25 & 2460774.9076 & 1.18 & 368 & 5.1 & $-1.03\pm2.62$ & $0.352\pm0.069$ & $-4.377\pm0.116$ & $0.318\pm0.002$ & $0.854\pm0.003$ \\
Apr 09 & 11:04:47 & 2460774.9669 & 1.20 & 345 & 5.6 & $-1.65\pm2.85$ & $0.345\pm0.079$ & $-4.389\pm0.136$ & $0.319\pm0.002$ & $0.855\pm0.003$ \\
\bottomrule                                
\end{tabular}
\tablefoot{The columns are the following: 1) date of the observation, 2) universal time of the observation, 3) heliocentric Julian date, 4) rotational cycle as computed from Eq.~\ref{eq:ephemeris}, 5) signal-to-noise ratio per resolution element per polarimetric sequence at 650~nm, 6) RMS noise level of Stokes~$V$ signal in units of unpolarised continuum, 7) longitudinal magnetic field, 8) $S$-index, 9) $\log R'_\mathrm{HK}$ index, 10) H$\alpha$ index, and 11) \ion{Ca}{II} infrared triplet index. The first observation marked with * was not used due to low signal-to-noise ratio.}
\end{sidewaystable*}

\section{ZDI fit of Stokes~$V$ LSD profiles}\label{app:zdi_fits}

We fitted the observed Stokes~$V$ LSD profiles down to $\chi^2_r=1.26$, from an initial value of 2.0 which corresponds to a featureless magnetic map. In this ZDI reconstruction, we removed the observation on April 5th at 10:43:38 UT since the S/N was visibly lower than the other observations, which allowed us to improve the target $\chi^2_r$. 

The deviation of the $\chi^2_r$ from 1.0 may come from the unaccounted intrinsic evolution of magnetic features, as well as the assumption of solid body rotation. \citet{Donahue1996} measured P$_\mathrm{rot}$ values for GJ~504 between 3.23\,d and 3.41\,d, and attributed this variation to differential rotation. The fact that we cannot constrain such phenomenon from our GJ~504 data does not imply a null surface shear. Rather, the limiting factor is likely the time span of our observations, which is too short for the surface latitudinal shear to have distorted the magnetic features enough to be detected.

\begin{figure}[!t]
\centering
    \includegraphics[width=0.5\columnwidth]{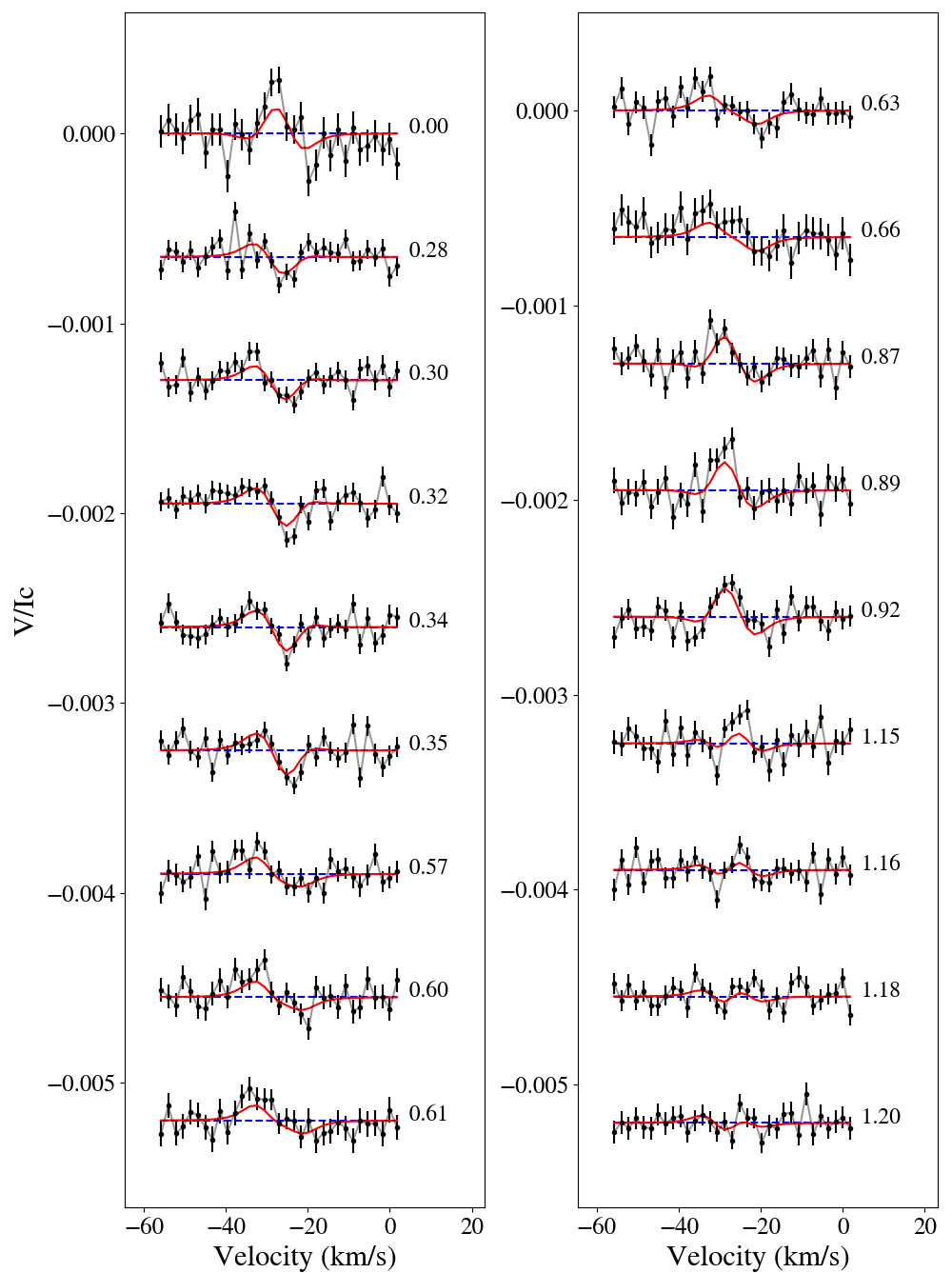}
    \caption{Time series of Stokes~$V$ LSD profiles and the ZDI models for GJ~504. The observations are shown in black and the models in red. The numbers on the right indicate the rotational cycle computed from Eq.~\ref{eq:ephemeris}. The horizontal dashed line represents the zero point of the profiles, which is shifted vertically based on the rotational phase for visualisation purposes. }
    \label{fig:stokesV}
\end{figure}

\section{Rotational evolution model setup}\label{app:rot_model}

To compute the rotational and X-ray luminosity tracks of GJ~504 for the evolutionary scenarios with and without engulfment, we provided the structural tracks computed with the CLES stellar evolution code \citep{Scuflaire2008} to a SPI code \citep{Privitera2016b, Rao2018, Pezzotti2021, Pezzotti2025}. In this code, the treatment of two main SPIs is implemented: the gravitational-tidal one, accounting for the exchange of angular momentum between the host star's surface and the planetary orbit, and the radiative one, through which the mass loss from planetary atmosphere due to the host star high energy radiation is estimated.

For the evolution of the host star surface rotation rate, the impact of magnetised winds is considered by means of the solar calibrated prescription of \citet{Matt2015, Matt2019}. Solid body rotation is assumed for GJ~504 on the pre-main sequence and main sequence phases \citep{Rao2021}. The evolution of the X-ray luminosity is consistently computed with respect to the one of the surface rotation rate, by using the recalibrated $\rm R_{\mathrm{x}}-R_{\mathrm{o}}$ relationship of \citet{Johnstone2021}, as in \citet{Pezzotti2021}, where $\rm R_{\mathrm{x}}$ is the ratio between the X-ray and the bolometric luminosity, and $\rm Ro$ is the stellar Rossby number.

\section{Scaling the torque}\label{app:torque_scaling}

The normalisation factors used to reconcile the angular momentum loss estimated from ZDI maps with the ones from rotational models depend on the considered braking law and modelling assumptions \citep[see e.g.][]{Evensberget2024}. In \citet{Finley2019a}, the authors found that torque estimates based on rotation-evolution models, as in \citet{Matt2015} are systematically larger than the ones derived from ZDI-based wind models \citep[e.g.][]{Finley2018}. They suggest several causes at the source of this discrepancy, among which potential issues with the rotation-evolution models or systematic effects from the ZDI technique in underestimating field strengths. 

In \citet{Finley2019a}, they show that multiplying the $\rm \dot{J}$ estimates by 20, for their sample of stars with ZDI-based wind models, a general better overlap is obtained with respect to the torques estimated from rotation-evolution models. By comparing rotational tracks computed for a standard solar model, with the same method and braking law as in \citet{Pezzotti2025}, we found that a factor of 20 is indeed needed to reconcile the rotational tracks at at the solar age with the torque determined at the solar maximum. For consistency, in this work we multiplied the $\mathrm{\dot{J}}$ derived in Sect.\ref{sec:wind} by a factor 20 to reconcile the predictions from theoretical rotational tracks with the semi-empirical estimates of the torque. In the $\rm \dot{J}_{wind}$ panel of Fig.~\ref{Fig:single_all}, we indicated the angular momentum loss we derived in Sect.~\ref{sec:wind} multiplied by 20 ($\rm \dot{J}_{wind, ZDI} \times 20 = 1.51 \times 10^{33}~erg$). We also note that the minimum amount of scaling for $\rm \dot{J}_{wind, ZDI}$ to be compatible with the engulfment scenario (precisely, the evolutionary track of $3.2\Omega_\odot$) is a factor of approximately four.

Owing to the fact that ZDI recovers only a fraction of the total magnetic field strength \citep[see e.g.][]{Yadav2015}, we decided to perform a simulation of the stellar wind with augmented magnetic field strength. In practice, we followed \citet{Evensberget2023} and multiplied the input radial magnetic field strength of ZDI by a factor of five, and then characterised the wind properties in a similar manner to Sect.~\ref{sec:wind}. We obtained $\rm \dot{J}_{wind, B \times 5} = 7.15 \times 10^{32}~erg$. As shown in Fig.~\ref{Fig:single_all}, both the rescaled values of $\rm \dot{J}$ are compatible with the planetary engulfment scenario.

\section{Comparative analysis with HD~75332}\label{app:HD75332}

GJ~504 fundamental parameters are close to $\rm \tau~ Boo$ ($\rm M = 1.39~M_\odot$, $\rm P_{rot}= 3.1\,d$, $\rm age = 1.9~Gyr$) and HD~75332 ($\rm M =1.21~M_\odot$, $\rm P_{rot}= 3.56\,d$, $\rm age = 0.9~Gyr$), as can be seen in Table~\ref{tab:star_properties}. In this appendix, we present a similar analysis for HD~75332 to that described in Sect.~\ref{sec:rot_models}, using the magnetic properties provided in \citet{Brown2021}, in order to assess whether it could have undergone an evolutionary history similar to GJ~504. $\rm \tau~ Boo$ is excluded from this comparison because the star is known to host a hot Jupiter planet \citep{Butler1997}, hence the engulfment scenario of a tidally interacting hot Jupiter would render the analysis inconsistent.

We searched for the optimal fundamental stellar parameters of HD~75332 by means of a global minimisation technique carried out with SPInS \citep{Lebreton2020}, on the basis of a grid of stellar tracks computed with the CLES stellar evolution code \citep{Scuflaire2008}. We used $\rm T_{eff}[K] = 6258 \pm 44$, $\rm log~g = 4.34 \pm 0.03$ and $\rm [M/H] = 0.05 \pm 0.03$ from \citet{Brown2021}, and the bolometric luminosity derived from the Gaia DR3 \citep{Gaia2020} G-band ($\rm L_{bol}[L_{\odot}] = 1.99 \pm 0.01$) as in \citet{Buldgen2019}, as constraints for the modelling. The optimal values that we found are: $\rm M = 1.15 \pm 0.06~M_{\odot}$, $\rm R = 1.19 \pm 0.03~R_{\odot}$, $\rm Age = 1.78 \pm 0.68~Gyr$. 

\begin{figure}[!b]
    \centering
    \includegraphics[width=\textwidth]{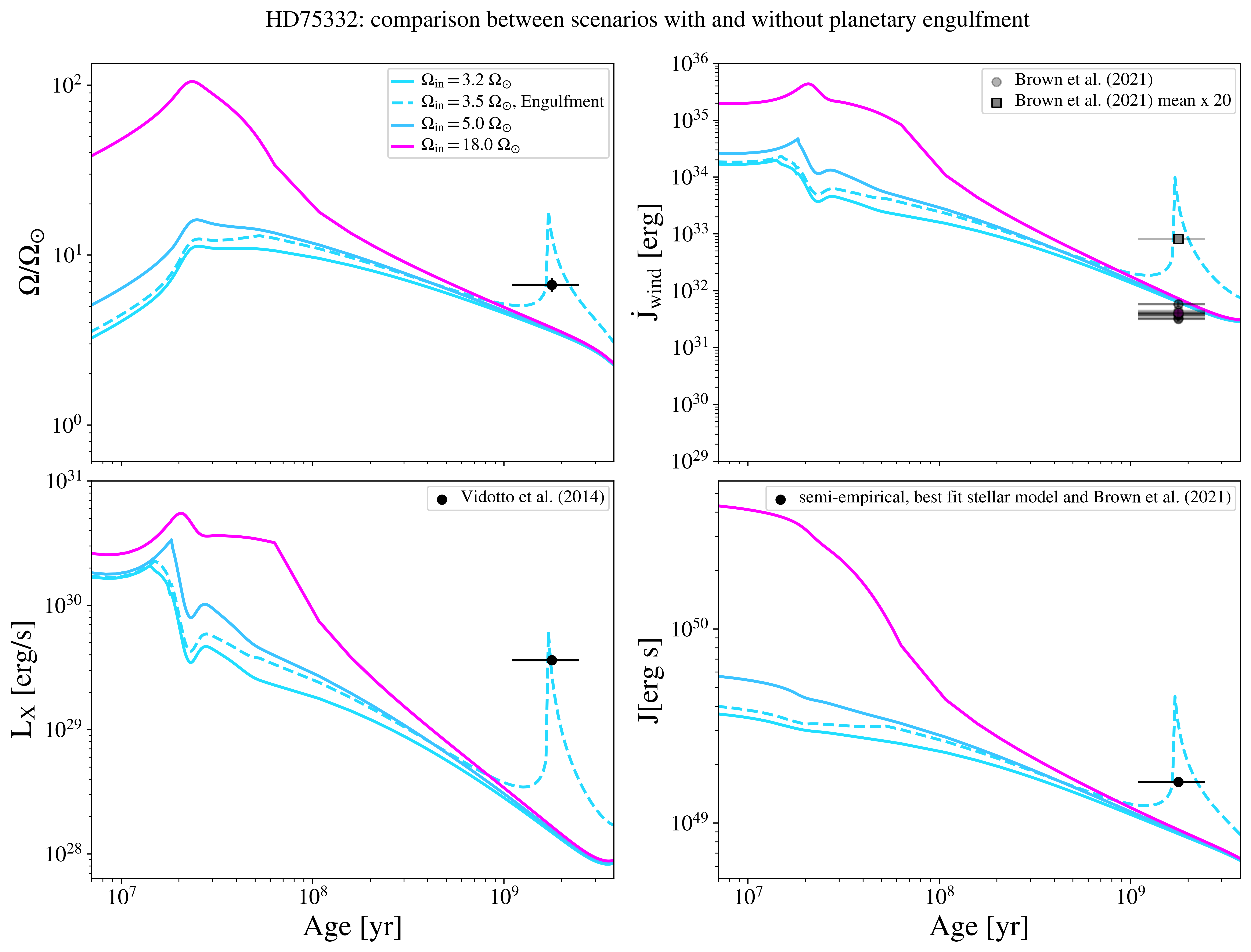} 
    \caption{Evolutionary tracks computed for HD~75332. The scenarios without (solid lines) and with (dashed line) the engulfment of a $\rm 3~M_{J}$ planet at an initial orbital distance $\rm a_{in} = 0.025~AU$ are included. For the stellar initial surface rotation rate, we considered values of $\rm \Omega_{in}(\Omega_{\odot}) = 3.2, 5.0, 18$, as representative of the evolution as slow, moderate and fast rotator \citep{Eggenberger2019a}. The spike featured by the dashed line indicates the effect the planetary engulfment has on the stellar property. In the top-left panel, the black dot shows the observed surface rotation rate as in \cite{Brown2021}. In the top-right panel, the black dots shows the estimates of angular momentum loss derived by using the prescription in \cite{Finley2018} and the values in Table~2 of \cite{Brown2021}. The gray square shows the mean value of the angular momentum loss estimates multiplied by a factor 20. In the bottom-left panel, the back dot shows the observational value of the X-ray luminosity from \cite{Vidotto2014a}. In the bottom-right panel, the black dot shows the semi-empirical angular momentum obtained by multiplying the best-fit stellar model momentum of inertia by the surface rotation rate in \cite{Brown2021}.}
    \label{Fig:HD75332}
\end{figure}

We then computed evolutionary tracks for HD~75332 as shown in Fig.~\ref{Fig:HD75332}. We assumed the same two scenarios as for GJ~504: evolution without planet, and with an inward migration of a giant planet. In the scenario without planet, we considered three initial surface rotation rates $\rm \Omega_{in} = 3.2, 5, 18~\Omega_{\odot}$ and in the scenario with a planet, we considered a $\rm 3~M_{J}$ companion at initial orbital distance $\rm a_{in}(AU) = 0.025$, and host star initial surface rotation rate $\rm \Omega_{in}(\Omega_{\odot}) = 3.5$. We selected this value for the initial surface rotation rate for analogy with the case of GJ~504, for which the engulfment and consequent compatibility with the observational and semi-empirical constraints for $\rm \Omega$, $\rm L_X$, $\rm \dot{J}_{wind}$ and $\rm J$ is obtained for models with $\rm \Omega_{in}(\Omega_{\odot}) \sim 3.2 - 4$. In this scenario, the planet is engulfed by the host star at $\rm \sim 1.5~Gyr$. It is worth recalling that the choice of initial mass orbital distance for the planet, together with the initial surface rotation rate for the host star, is degenerate. Therefore, choosing a different set of initial parameters may lead to a similar result, as discussed in \cite{Pezzotti2025}.

In Fig.~\ref{Fig:HD75332}, we show the comparison between observational constraints and the evolutionary tracks for $\rm \Omega_{in}$, angular momentum loss, $\rm L_X$, and total angular momentum. The observational constraints of $\rm \Omega_{in}$ and $\rm L_X$ were taken from \cite{Brown2021} and \citet{Vidotto2014a}, respectively. For the angular momentum loss, the black circle indicates the value computed using the prescription of \citet{Finley2018}. The dipolar, quadrupolar and octupolar components of the large-scale poloidal magnetic field of HD~75332 were taken from Table~3 in \citet{Brown2021}. In Fig.~\ref{Fig:HD75332}, the solid lines indicate the scenario without planet engulfment, while the dashed line represents the track relative to the engulfment scenario. We find that all the observational constraints (stellar rotation, angular momentum, X-ray luminosity, and total angular momentum) are reconciled with the theoretical tracks only with the engulfment scenario, in a similar manner as for GJ~504. Analogously, the observed angular momentum loss is consistent with the evolutionary track associated with the planetary engulfment when multiplied by a factor of 20.

From this comparative analysis, we infer that GJ~504 and HD~75332 may have experienced a similar evolutionary history, potentially involving the engulfment of a giant planet. However, several uncertainties persist regarding the magnetic properties of stars occupying this particular region of the mass–period parameter space. In other words, the magnetic properties of only GJ~504, HD~75322 and $\rm\tau~Boo$ have been reconstructed in the literature. \citet{Pezzotti2025b} investigate the activity–rotation–age relation for 13 G-type and F-type stars with high-quality asteroseismic constraints from the \textit{Kepler} LEGACY sample. The authors report that none of these $Kepler$ stars, despite having highly accurate and precise fundamental parameters like rotation periods and eROSITA X-ray luminosities, exhibit anomalous properties similar to those of GJ~504 and HD~75332. Specifically, the rotation periods are not substantially shorter than model predictions and the X-ray luminosities are not enhanced given their stellar age.

\end{appendix}

\end{document}